\documentclass[submission,copyright,creativecommons]{eptcs}
\providecommand{\event}{SCSS 2021} % Name of the event you are submitting to
\pdfoutput=1
\usepackage{underscore}           % Only needed if you use pdflatex.
\usepackage{times}
\usepackage{graphicx}
\usepackage{latexsym}
\usepackage{amsmath}
\usepackage[utf8x]{inputenc}
\usepackage[frozencache=true,cachedir=minted-cache]{minted}
\usepackage{caption}
\usepackage{xspace}
\usepackage{tikz}
\usepackage{pgfplots}

\newcommand{\argot}{\texttt{ArGoT}\xspace}
\newcommand{\isa}{\texttt{IS-A}\xspace}

\title{\texttt{ArGoT}: A Glossary of Terms extracted from the arXiv}
\author{Luis Berlioz
    \institute{University of Pittsburgh\\
    Pennsylvania, USA}
\email{lab232@pitt.edu}
}
\def\titlerunning{\texttt{ArGoT}: A Glossary Terms extracted from the arXiv}
\def\authorrunning{Luis Berlioz}
\begin{document}
\maketitle

\begin{abstract}
    We introduce ArGoT, a data set of mathematical terms extracted from 
the articles hosted on the arXiv website. 
A term is any mathematical concept defined in an article.
Using labels in the article's source code and examples from other
popular math websites, we mine all the terms in the
arXiv data and compile a comprehensive vocabulary of mathematical terms.
Each term can be then organized in a dependency graph by using
the term's definitions and the arXiv's metadata.
Using both hyperbolic and standard word embeddings, we demonstrate how 
this structure is reflected in the text's vector representation and
how they capture relations of entailment in mathematical concepts. This
data set is part of an ongoing effort to align natural mathematical
text with existing Interactive Theorem Prover  Libraries (ITPs) of
formally verified statements.
%% Include Formal abstracts project

\end{abstract}

\section{Introduction and Motivation}
Mathematical writing usually adheres to strict conventions of rigor
and consistent usage of terminology.
New concepts are usually introduced in characteristically worded 
definitions (with patterns like \textit{if
    and only if} or \textit{we say a group is abelian...}). 
This feature can be used to train language models to detect if a term is defined in a text.
%Also, the old 
%terms on which the new ones depend are seldom skipped.  
Using this, we have created \argot (\textbf{ar}Xiv \textbf{G}lossary \textbf{o}f \textbf{T}erms), a silver standard data set of terms defined in
the Mathematical articles of the arXiv website. 
We showcase several interesting applications of this data. The data set includes the articles and  paragraph number in which each term appears. By using article metadata, we show that this can be an effective way of assigning an arXiv mathematical category\footnote{arXiv's categories within mathematics: \url{https://arxiv.org/archive/math}} to each term.
%We demonstrate that hyperbolic word embedding language models  
%can effectively capture relations of entailment in advanced
Another application is to join the terms with more than one word into a single token.
These phrases usually represent important mathematical concepts with a specific meaning.
We show how standard word embedding models like word2vec
\cite{word2vec} and GloVe \cite{DBLP:conf/emnlp/PenningtonSM14} capture this by embedding phrases instead of individual words.
Even more, the word-vector can be used to predict which mathematical field the term belongs to, and hypernimity relations.
%And in the future, will provide abundant training examples for NLU and automated reasoning.

All these properties  makes \argot  a data set that will be of
interest to the broader NLP research community by providing abundant
examples for automated reasoning and NLU systems. 
Our main objective is to organize a comprehensive dependency graph of mathematical concepts that can be aligned with existing libraries of formalized mathematics like \texttt{mathlib}.\footnote{https://github.com/leanprover-community/mathlib}
The data is downloadable from \url{https://sigmathling.kwarc.info/resources/argot-dataset-2021/} and the all the code that went
into producing it is in: \url{https://github.com/lab156/arxivDownload}
%the text with a unique predictability that enables the NLP researcher to perform
%adventurous experimentation and simple debugging.

This data set was created  as part of the Formal Abstracts project.
Our group has benefited from a grant from the Sloan  Foundation
(G-2018-10067) and from the computing resources startup allocation
\mbox{\#TG-DMS190028} and \mbox{\#TG-DMS200030} on the Bridges-2
supercomputer at the Pittsburgh Supercomputing Center (PSC).
\begin{table}[h]
    \centering
    \begin{minipage}{0.45\textwidth}
        %\centering
%\begin{table}
    \small
\centering
\begin{tabular}{lr}
    \hline \textbf{Term} &  \textbf{Count} \\ \hline
lie algebra & 20524 \\
%suppose & 18043 \\
hilbert space & 16881 \\
function & 14920 \\
banach space & 14461 \\
metric space & 12882 \\
\_inline\_math\_-module & 12731 \\
topological space & 12518 \\
%sequence & 12308 \\
disjoint union & 11436 \\
vector space & 11337 \\
simplicial complex & 10943 \\
%graph & 10811 \\
%map & 10654 \\
%morphism & 10596 \\
\hline

\end{tabular}
\caption{\label{term-cnt-tab} Most common multiword entries in the data base. }
%\end{table}
        \end{minipage}\hfill
    \begin{minipage}{0.45\textwidth}
%\begin{table}
    \small
\centering
\begin{tabular}{lrrr}
    \hline
    \multicolumn{4}{c}{Classification Task} \\
    \hline
\textbf{Method}  & \textbf{Precision} &  \textbf{Recall} & \textbf{F1}\\ 
\hline
    SGD-SVM & 0.88 & 0.87 & 0.87 \\
    Conv1D & 0.92 & 0.92 & 0.92 \\
    BiLSTM & 0.93 & 0.93 & 0.93 \\
     \hline
    \hline
    \multicolumn{4}{c}{NER Task} \\
    \hline
    ChunkParse & 0.32 & 0.68 & 0.43 \\
    LSTM-CRF & 0.69 & 0.65 & 0.67 \\
    \hline
\end{tabular}
\caption{\label{metric-comp} Training metrics on the classification and NER tasks.}
%\end{table}%% End of second table
\end{minipage}
\end{table}

\section{Description of the Term-Definition Extraction Method}
In \cite{DBLP:conf/mkm/Berlioz19, Deyan1}, the authors describe the 
method used  to obtain the training data for a text classification
model that identifies definitions and the Named Entity Recognition (NER) model that identifies the term being defined. 

The classification task consists of training a binary classifier to determine whether a paragraph is a definition or not. We use the \verb/\begin{definition}...\end{definition}/ in the article's \LaTeX{} source to identify true examples. To gather non-definitions, we randomly sample paragraphs out of the same articles.
The source of the training data is the \LaTeX{} source code of the articles available from the arXiv website. A total of 1,552,268 paragraphs labeled as definitions or non-definitions were produced for training. It was split as follows: 80\% training 10\% testing and 10\% validation. This data was used to train three different and common classification models:
\begin{itemize}
    \item The Stochastic Gradient Descent with Support Vector Machines (SGD-SVM). 
    \item The one-dimensional convolutions (Conv1D) neural network.
    \item And Bidirectional LSTM (BiLSTM). 
\end{itemize}
For the first method, we used the implementation distributed with scikit-learn library \cite{DBLP:journals/jmlr/PedregosaVGMTGBPWDVPCBPD11}. The last two were implemented in Tensorflow. Table \ref{metric-comp} shows the most common metrics of performance for each method. 

The definitions are then fed into a NER model to identify the term being defined in them.
The data used to train the NER model comes from the Wikipedia English dump\footnote{\url{https://dumps.wikimedia.org/}} and several mathematical websites like
PlanetMath\footnote{\url{https://planetmath.org/}} and The Stacks
Project.\footnote{\url{https://stacks.math.columbia.edu/}}

We tested two different implementations of the NER system, the first is the \emph{ChunkParse} algorithm available from the NLTK library \cite{DBLP:books/daglib/0022921}. The second is a time-distributed LSTM (LSTM-CRF) \cite{DBLP:journals/corr/HuangXY15}. Both architectures use a similar set of features that in addition to the words that form the text, detect if the word is capitalized, its part-of-speech (POS) and parses punctuation e.g. to tell if a period is part of an abbreviation or an end of line.
To compare the two implementations, we used the \texttt{ChunkScore} method in
the NLTK library \cite{DBLP:books/daglib/0022921}. The results appear in Table \ref{metric-comp}. 

We have compiled two different and independent glossaries by running the 
algorithm through all of the arXiv's mathematical content. 
The first one  is based on neural networks (NN), it uses  LSTM for both the classification and NER tasks. 
In contrast, the second one combines the SGD and ChunkParser method to provide a completely independent approach to the previous model.

It is interesting to compare the results obtained using the two models. For the classification task, we have observed Cohen's kappa ($\kappa$) inter-rater agreement of 93\% between the results produced by the two methods.
This corresponds to a high degree of agreement between the two classifiers \cite{cohenkappa}.

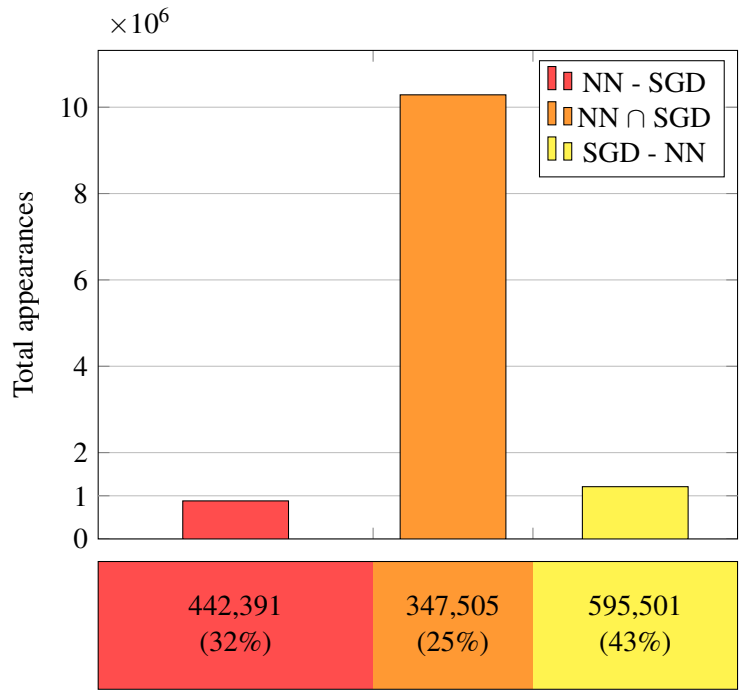
\begin{figure}
    \centering
    \begin{tikzpicture}[scale=1]
\def\mainwidth{8.5}
\def\MW{\mainwidth}
\pgfplotsset{width=\mainwidth cm, height=6.5cm, compat=1.3}

\def\NNx{.43}
\def\NN{\NNx*\mainwidth}

\def\Intx{0.25}
\def\Int{\Intx*\mainwidth}

\def\SGDx{0.32}
\def\SGD{\SGDx*\mainwidth}

\def\boxHeighta{-0.3}
\def\boxHeightb{-2}
\def\boxcc{\boxHeighta/2+\boxHeightb/2}

\tikzset{NN style/.style={ fill=red!70, mark=none} }
\tikzset{Int style/.style={ fill=orange!80, mark=none} }
\tikzset{SGD style/.style={ fill=yellow!80, mark=none} }

\begin{axis}[
xmax=1.0, xmin=0,
ymin=0.0, 
scale only axis,
bar width=40,
bar shift=0,
ybar legend,
xtick={ \NNx, \NNx+\Intx, \NNx+\Intx+\SGDx },
xticklabels={ , , },
extra y ticks={1000000.0},
extra y tick labels={1},
ymajorgrids,
%scaled y ticks=manual:{$\times 10^6$}{1000000},
scaled y ticks=real:1000000,
ytick scale label code/.code={$\times 10^6$},
%title=Unique terms vrs. Terms with repetitions,
ylabel={Total appearances},
]
\addplot+[ybar, style=NN style, draw=black] plot coordinates { (\NNx/2,881616) };
\addplot+[ybar, style=Int style, draw=black] plot coordinates { (\NNx +\Intx/2,10286789) };
\addplot+[ybar, style=SGD style, draw=black] plot coordinates { (\SGDx/2+\Intx+ \NNx,1210995) };

\legend{NN - SGD,NN $\cap$ SGD, SGD - NN}

\end{axis}
\draw[style=NN style, draw opacity=0.0] (0,\boxHeighta) rectangle (\NN, \boxHeightb);
\draw[style=Int style, draw opacity=0.0] (\NN, \boxHeighta) rectangle (\NN+\Int, \boxHeightb);
\draw[style=SGD style, draw opacity=0.0] (\NN+\Int, \boxHeighta) rectangle (\SGD+\Int+\NN, \boxHeightb);
\draw[draw] (0, \boxHeighta) rectangle (\SGD+\Int+\NN, \boxHeightb);

\node[align=center] (NNc) at (\NN/2, \boxcc) {442,391\\(32\%)};
\node[align=center] (Intc) at (\NN+\Int/2, \boxcc) {347,505\\ (25\%)} ;
\node[align=center] (SGDc) at (\NN+\Int+\SGD/2, \boxcc) {595,501\\ (43\%)};

\end{tikzpicture}
    \caption{\label{sizes} Comparison of the two glossaries. The bar graph on top counts the total appearances of a term in both the NN and SGD glossaries. The bottom compares the relative sizes of the NN-only, intersection, and SGD-only distinct terms.}
\end{figure}

As for the final results, Figure \ref{sizes} compares the two glossaries by counting the number of times a term appears in either glossary, and the number of distinct terms.
The results point to a high consistency of the two systems on a relatively small set of 350,000 terms.

Table~\ref{term-cnt-tab} lists some of the most frequently found terms in the
data set. 

\subsection{Format and Design of the Data Set}
The \argot data set is distributed in the form of compressed
XML
files that follow the same naming convention the arXiv's bulk download
distribution.\footnote{arXiv Bulk Data Access: \url{https://arxiv.org/help/bulk\_data}}
For instance, Table \ref{glossary-example} shows a sample entry
in the fifth file corresponding to July, 2014. The definition's
statement and terms (definiendum) are specified in the \texttt{stmnt} and
\texttt{dfndum} tags respectively and the paragraph \texttt{index} is
specified as an attribute of the \texttt{definition} tag.

%%% database entry example
\begin{table*}[h]
    \centering
    \begin{minted}[fontsize=\small]{xml}
    <article name="1407_005/1407.2218/1407.2218.xml" num="89">
    <definition index="51">
        <stmnt> Assume _inline_math_. We define the following space-time 
        norm if _inline_math_ is a time interval _display_math_ </stmnt>
        <dfndum>space-time norm</dfndum>
    </definition>
    </article>
\end{minted}
\caption{\label{glossary-example} Example of an entry in the term's data set. The statement of the definition is contained in the $<$stmnt$>$ tag. The terms (definiendum) are listed as $<$dfndum$>$ tags. Each entry contains all the information to recover, article's name and paragraph's position.}
\end{table*}

%\begin{table}
\begin{figure}[h!]
\centering
\begin{minipage}{0.4\textwidth}
    \footnotesize
\begin{tabular}{lc}
\hline
\textbf{Category:} & \textbf{Count}\\
\hline
math.FA& 5922 \\
 math.AP& 2045 \\
 math.PR& 1022 \\
 math.DS& 833 \\
 math.OA& 595 \\
 math.CA& 535 \\
 math.DG& 483 \\
 math-ph& 466 \\
 math.OC& 398 \\
 math.CV& 304 \\
 math.NA& 275 \\
 math.GR& 226 \\
 math.MG& 173 \\
 math.LO& 168 \\
 math.SP& 163 \\
math.NT& 131 \\
\hline
\end{tabular}
\begin{tabular}{lc}
\hline
\textbf{Category:} & \textbf{Count}\\
\hline
 math.GN& 108 \\
 math.RT& 85 \\
 math.SG& 77 \\
 math.GT& 76 \\
 math.CO& 61 \\
 math.ST& 61 \\
 math.KT& 50 \\
 math.GM& 48 \\
 math.AG& 35 \\
 math.RA& 33 \\
 math.HO& 32 \\
 math.CT& 23 \\
 math.AT& 15 \\
 math.QA& 10 \\
 math.AC& 8 \\
         & \\
\hline
\end{tabular}
\captionof{table}{Category profile for the term: \emph{Banach Space}. The codes
    are part of the metadata for each arXiv article.
}\label{tab:categories}
%\end{table}
\end{minipage}\hfill
\begin{minipage}{0.55\textwidth}
%\begin{figure}[ht]
    \centering
    \includegraphics[width=0.99\textwidth]{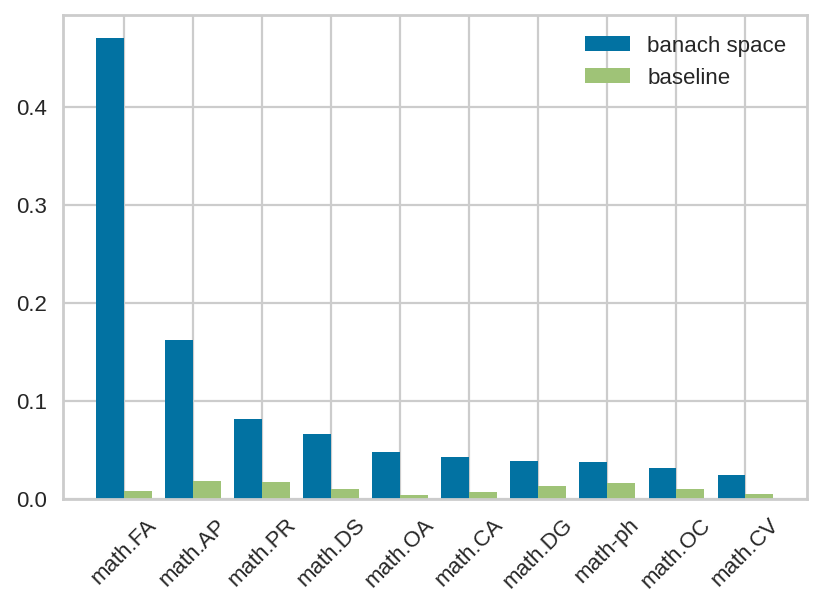}
    \captionof{figure}{\label{bar} Comparison between the term's category
        distribution and baseline distribution. Only categories with
        the highest values for the term are shown.}
\end{minipage}
\end{figure}

%\subsection{Related Work}
%There have been several recent projects related to NLP and information extraction from scholarly papers. In \cite{kang-etal-2020-document}, the 

\section{Augmenting Terms with arXiv's Metadata}
Each mathematical article in the arXiv is classified in one or more
\emph{categories}  by the author at
the time of submission. Categories include \texttt{math.FA} and  \texttt{math.PR} which stand for Functional Analysis and Probability respectively. The full list is available at \url{https://arxiv.org/archive/math}.
This is part of  the arXiv's metadata and also records information like the list of authors, math subject classification (MSC)~codes, date of submission, etc. 

By counting the categories in which a certain term is used, we get an
idea of the subjects that it belongs to. In Table
\ref{tab:categories}, we see the category profile of a very common
term. Since the number of articles in each category varies
significantly, we also take into account the baseline distribution,
that is, the ratio of articles in each category to the total number of
articles.
Hence, it is possible to give an empirical score of a term's
pertinence to a certain category by comparing its category profile
with the baseline distribution. In order to measure how much of an 
outlier a term is to the baseline distribution, we use the KL-divergence:
$$D_{\text{KL}}(P \Vert Q) = \sum_{x\in X} P(x)\log(P(x)/Q(x)),$$
where $P$ and $Q$ are the probability distributions of the term and the baseline respectively. And, $X$ is the set of all the categories.

%of the math articles in the following ways: First, We remove all \LaTeX{} specific code, leaving only the natural language text. Second, we normalize the text. This includes converting all characters to lowercase, removing accents from letters, removing non-ascii characters, etc. Lastly, we 

\begin{figure*}
    \centering
    \includegraphics[width=0.82\textwidth]{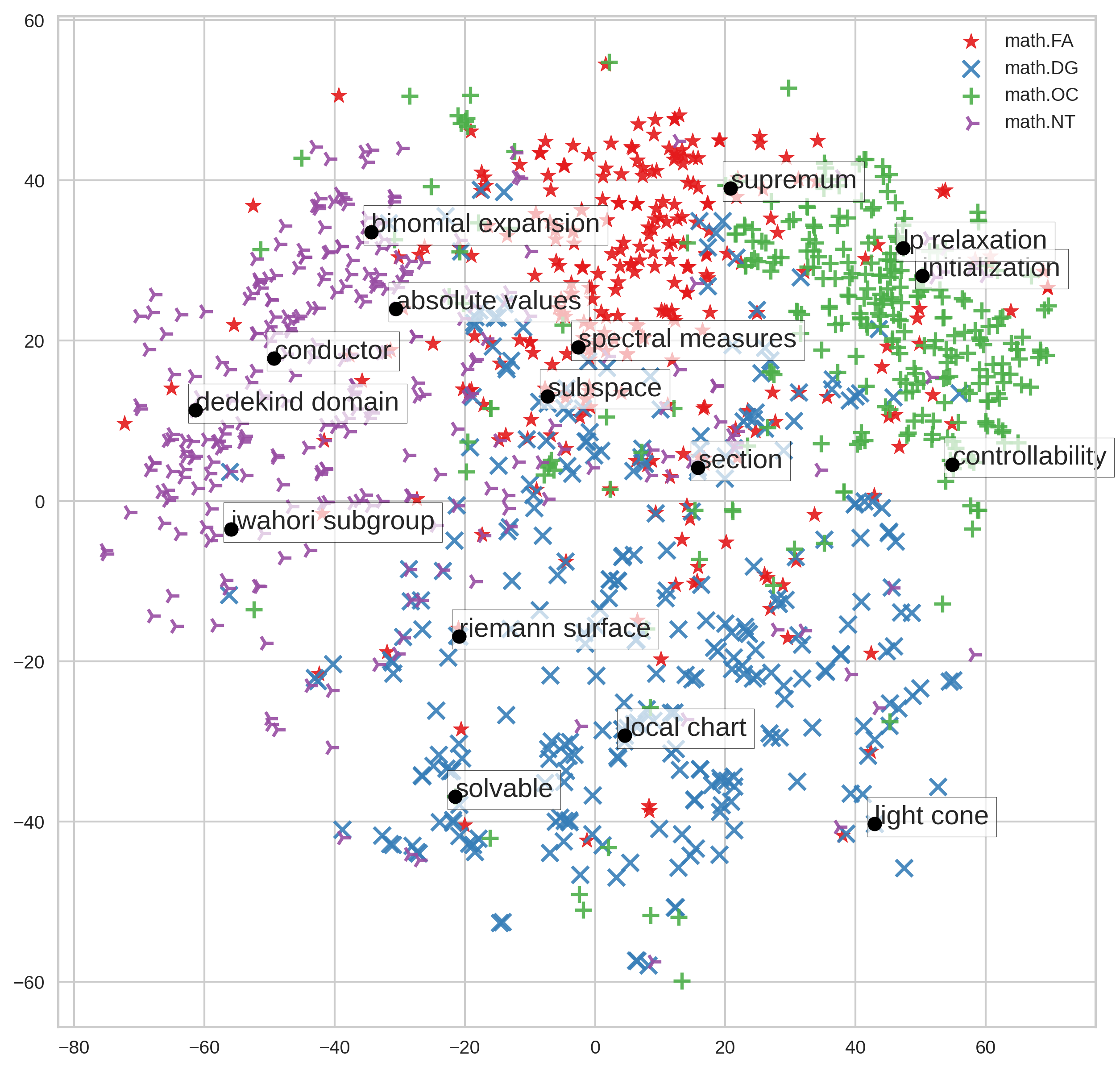}
    \caption{\label{scatter} t-SNE visualization of the word vectors
        of selected terms in the data set. The terms are selected to be specific to 
        the four categories in the picture.  Points with a label are selected
    at random.}
\end{figure*}

The next step is to generate word embeddings. To prepare for this, we modify the text by joining multiword terms in \argot to produce individuals tokens. After normalizing the text, i.e. converting to lowercase and removing punctuation and special characters; the result is a large amount of text that is ready to be consumed by either the word2vec or GloVe algorithms. 
In Figure~\ref{scatter}, we observe a t-SNE (t-distributed stochastic
neighbor embedding) visualization of a word2vec 
model produced this way. In this image, each term is assigned its most frequent category. Notice that even though the \argot data set has no access to the arXiv categories, the vectors in the same category cluster together.
We consider this as a strong indication of alignment between clusters and categories.

%\section{Distributed Representations and Semantic Evaluation}
%\section{Distributed Representations with the arXiv's Content} 
%We produced word embeddings of the arXiv in two rounds. The first one
%was used in the classification and NER tasks, and the second one in
%the embedding of joined terms. The source was processed in the
%following ways: Substitute all math formulas, citations and
%references with tokens, i.e. (\_inline\_math\_, \_cite\_, \_ref\_).
% Perform usual text normalization and tokenization for word 
%embeddings. This includes removal of caps, numbers, non-ascii
%characters, etc. In the second round,  Join the occurring instances 
%of a term to create a unique token, for instance: \emph{banach space}
%is joined as \emph{banach\textbf{\_}space}.
%We produced both GloVe and word2vec word
%embeddings and noticed no significant difference in performance.

\section{Using Hyperbolic Word Embeddings to Extract Hypernymy Relations}
It is natural to want to organize mathematical concepts into taxonomies of various sorts.
For instance, the SMGloM project \cite{DBLP:conf/icms/GinevIJKKOSSTW16} introduced a rich standard for mathematical ontologies. Another approach aims to create a semantic hierarchy of concepts such that for a given term we can enumerate all its hypernyms \cite{wang-etal-2017-short}.

This can be  achieved by counting the co-occurrence \cite{hearst-1992-automatic} of terms in definitions. This approach has certain drawbacks, for instance, it relies on co-occurrence examples for each pair of terms, this ends up producing an abundance of disconnected (i.e. not co-occurring) terms \cite{aly-etal-2019-every}.

Another possibility, involves the use of \emph{hyperbolic word embeddings}, in this setting the hypernimity relation becomes a geometric vector in hyperbolic space. This implies that every two terms in the embedding can be compared by using the hyperbolic metric.
This type of word embeddings is known to outperform euclidean models in the representation of hierarchical structures \cite{DBLP:conf/nips/NickelK17}.

We used PoincareGlove \cite{DBLP:conf/iclr/TifreaBG19} to create hyperbolic
word embeddings. This algorithm modifies the GloVe \textit{euclidean} objective function to use a hyperbolic metric instead. 
In addition to the same text input as word2vec and GloVe, this model requires a small set of examples in order to interpret the embedding. 
For general purpose English text, WordNet \cite{wordnet} is the standard  choice. 
In WordNet, every entry is assigned an integer level in a hypernymy hierarchy (this is the
\texttt{max\_depth} attribute of the NLTK's WordNet
API).\footnote{\url{https://www.nltk.org/howto/wordnet.html}} 

To generate something  analogous to WordNet levels for mathematical content, we opted for the PlanetMath data set. This is due to its relatively small size, broad coverage of mathematical knowledge and independence of the arXiv data. 
 Given two term-definition pairs $(t_1, D_1)$ and $(t_2, D_2)$, we say that term
$t_2$ \emph{depends} on the term $t_1$ if $D_2$ contains $t_1$.
For small sets of term-definition pairs with no interdependence, this
simple criterion is enough to create a directed graph $(V, E)$ where
$V$ is the set of all the terms and $E$ is the set of all the
dependency relations. To assign a level $\lambda (v)$ to every 
vertex $v\in V$, solve the following integer linear program:
%\begin{align*}
%    \text{min} & \quad \sum_{(v,w) \in E} \lambda(w) - \lambda(v)  \\
%    \text{such that} & \quad \lambda(w) - \lambda(v) \geq 1  \\
%     & \forall (v,w) \in E. 
%\end{align*}
$$\min  \sum_{(v,w) \in E} \lambda(w) - \lambda(v),  \quad \text{such that}  \quad \lambda(w) - \lambda(v) \geq 1\quad \forall (v,w) \in E. $$

This linear model appears in \cite{DBLP:journals/tse/GansnerKNV93} as a subtask of a directed graph drawing algorithm. There, it is used to estimate the ideal number of levels to draw a directed graph.

Table \ref{tab:hypernyny} shows the nearest neighbors of four
different terms.  The neighbors are found using the Euclidean distance. 
The terms are 
sorted in order of the average value of their $y$-coordinates (which in
the upper-half plane model represents the variance of the underlying
Gaussian distribution). This is referred to as the \isa rating.

\begin{table}
    \small
\centering
\begin{tabular}{ll|ll}
    \hline \textbf{Term} &  \isa &  
    \textbf{Term} &  \isa \\ \hline
    hyperbolic\_metric & -1.11 &  &\\
euclidean\_metric & -0.59  & digraph & -0.51 \\
metrics & -0.58 & undirected\_graph & -0.35 \\
riemannian\_metric & -0.46  &  undirected & -0.20 \\
riemannian & -0.42  & \textbf{directed\_graph} &  0.0\\
riemannian\_manif & -0.40 & graph & 1.24 \\
curvature & -0.27  & & \\
\textbf{metric} & 0.0 & & \\
\hline
banach\_algebra & -1.11  & probability\_distr & -0.24 \\
normed\_space & -0.98 & \textbf{random\_variable} & 0.0 \\
banach\_spaces & -0.38 & expectation & 0.23 \\
banach & -0.29  & distribution & 0.46 \\
closed\_subspace & -0.25 & probability & 0.67 \\
\textbf{banach\_space} & 0.0 & & \\
norm & 0.79 & & \\

\end{tabular}
\caption{\label{tab:hypernyny} 
   Query results sorted by \isa score (terms in upper lines tend to depend semantically on lower lines).  Cosine similar words were sorted by the
    \isa rating of the term in bold font. }
\end{table}

\section{Conclusions and Further Work}
We introduced \argot, an comprehensive glossary of mathematics
automatically collected from the mathematical content on the arXiv website. 
Essentially, it is set of term-definition pairs, where 
each pair can be contextualized in a large semantic network of
mathematical knowledge, i.e., dependency graph. We also showed how this 
network is reflected in the latent space of its vector embeddings. This
has great potential for use in experimentation of natural language
algorithms, by providing a source of logically consistent data. 

This project is an ongoing effort to align mathematical concepts
in natural language with  online repositories of formalized
mathematics like
\texttt{mathlib}.\footnote{\url{https://github.com/leanprover-community/mathlib}} 
As described in \cite{DBLP:conf/cikm/KaliszykKMR16}, this type of alignment is called \emph{automatically found aligment}.

In the near future we plan to further improve on the classification 
and NER tasks by creating a data set using solely the neural version
of the classifier and NER model. Also,  by using state-of-the-art
methods like the masked transformer language model \cite{DBLP:conf/naacl/DevlinCLT19} to
further improve the results. 
We also plan to compile the complete dependency graph in one 
large graph database.

%This type of domain specific data collection 
%and ontology population is becoming more commonplace as NLP models
%improve in performance. We hope this data set and will be helpful 
%to the researchers a similar project of 

%In order to perceive this property, 
%It is common for mathematical terms to be composed of multiple tokens.
%For instance, \emph{Riemann integral} and
%\emph{integral domain}. This means that detecting the multi-word entitiesis  necessary in order to take full advantage of mathematical text. 

%\nocite{*}
\bibliographystyle{eptcs}
\bibliography{article}

\begin{thebibliography}{10}
\providecommand{\bibitemdeclare}[2]{}
\providecommand{\surnamestart}{}
\providecommand{\surnameend}{}
\providecommand{\urlprefix}{Available at }
\providecommand{\url}[1]{\texttt{#1}}
\providecommand{\href}[2]{\texttt{#2}}
\providecommand{\urlalt}[2]{\href{#1}{#2}}
\providecommand{\doi}[1]{doi:\urlalt{http://dx.doi.org/#1}{#1}}
\providecommand{\bibinfo}[2]{#2}

\bibitemdeclare{inproceedings}{aly-etal-2019-every}
\bibitem{aly-etal-2019-every}
\bibinfo{author}{Rami \surnamestart Aly\surnameend}, \bibinfo{author}{Shantanu
  \surnamestart Acharya\surnameend}, \bibinfo{author}{Alexander \surnamestart
  Ossa\surnameend}, \bibinfo{author}{Arne \surnamestart K{\"o}hn\surnameend},
  \bibinfo{author}{Chris \surnamestart Biemann\surnameend} \&
  \bibinfo{author}{Alexander \surnamestart Panchenko\surnameend}
  (\bibinfo{year}{2019}): \emph{\bibinfo{title}{Every Child Should Have
  Parents: A Taxonomy Refinement Algorithm Based on Hyperbolic Term
  Embeddings}}.
\newblock In: {\sl \bibinfo{booktitle}{Proceedings of the 57th Annual Meeting
  of the Association for Computational Linguistics}},
  \bibinfo{publisher}{Association for Computational Linguistics},
  \bibinfo{address}{Florence, Italy}, pp. \bibinfo{pages}{4811--4817},
  \doi{10.18653/v1/P19-1474}.

\bibitemdeclare{inproceedings}{DBLP:conf/mkm/Berlioz19}
\bibitem{DBLP:conf/mkm/Berlioz19}
\bibinfo{author}{Luis \surnamestart Berlioz\surnameend} (\bibinfo{year}{2019}):
  \emph{\bibinfo{title}{Creating a Database of Definitions From Large
  Mathematical Corpora}}.
\newblock In \bibinfo{editor}{Edwin~C. \surnamestart Brady\surnameend},
  \bibinfo{editor}{James~H. \surnamestart Davenport\surnameend},
  \bibinfo{editor}{William~M. \surnamestart Farmer\surnameend},
  \bibinfo{editor}{Cezary \surnamestart Kaliszyk\surnameend},
  \bibinfo{editor}{Andrea \surnamestart Kohlhase\surnameend},
  \bibinfo{editor}{Michael \surnamestart Kohlhase\surnameend},
  \bibinfo{editor}{Dennis \surnamestart M{\"{u}}ller\surnameend},
  \bibinfo{editor}{Karol \surnamestart Pak\surnameend} \&
  \bibinfo{editor}{Claudio~Sacerdoti \surnamestart Coen\surnameend}, editors:
  {\sl \bibinfo{booktitle}{Joint Proceedings of the {FMM} and {LML} Workshops,
  Doctoral Program and Work in Progress at the Conference on Intelligent
  Computer Mathematics 2019 co-located with the 12th Conference on Intelligent
  Computer Mathematics {(CICM} 2019), Prague, Czech Republic, July 8-12,
  2019}}, {\sl \bibinfo{series}{{CEUR} Workshop Proceedings}}
  \bibinfo{volume}{2634}, \bibinfo{publisher}{CEUR-WS.org}.
\newblock \urlprefix\url{http://ceur-ws.org/Vol-2634/WiP2.pdf}.

\bibitemdeclare{book}{DBLP:books/daglib/0022921}
\bibitem{DBLP:books/daglib/0022921}
\bibinfo{author}{Steven \surnamestart Bird\surnameend}, \bibinfo{author}{Ewan
  \surnamestart Klein\surnameend} \& \bibinfo{author}{Edward \surnamestart
  Loper\surnameend} (\bibinfo{year}{2009}): \emph{\bibinfo{title}{Natural
  Language Processing with Python}}.
\newblock \bibinfo{publisher}{O'Reilly}.
\newblock
  \urlprefix\url{http://www.oreilly.de/catalog/9780596516499/index.html}.

\bibitemdeclare{article}{cohenkappa}
\bibitem{cohenkappa}
\bibinfo{author}{Jacob \surnamestart Cohen\surnameend} (\bibinfo{year}{1960}):
  \emph{\bibinfo{title}{A Coefficient of Agreement for Nominal Scales}}.
\newblock {\sl \bibinfo{journal}{Educational and Psychological Measurement}}
  \bibinfo{volume}{20}(\bibinfo{number}{1}), pp. \bibinfo{pages}{37--46},
  \doi{10.1177/001316446002000104}.

\bibitemdeclare{inproceedings}{DBLP:conf/naacl/DevlinCLT19}
\bibitem{DBLP:conf/naacl/DevlinCLT19}
\bibinfo{author}{Jacob \surnamestart Devlin\surnameend},
  \bibinfo{author}{Ming{-}Wei \surnamestart Chang\surnameend},
  \bibinfo{author}{Kenton \surnamestart Lee\surnameend} \&
  \bibinfo{author}{Kristina \surnamestart Toutanova\surnameend}
  (\bibinfo{year}{2019}): \emph{\bibinfo{title}{{BERT:} Pre-training of Deep
  Bidirectional Transformers for Language Understanding}}.
\newblock In \bibinfo{editor}{Jill \surnamestart Burstein\surnameend},
  \bibinfo{editor}{Christy \surnamestart Doran\surnameend} \&
  \bibinfo{editor}{Thamar \surnamestart Solorio\surnameend}, editors: {\sl
  \bibinfo{booktitle}{Proceedings of the 2019 Conference of the North American
  Chapter of the Association for Computational Linguistics: Human Language
  Technologies, {NAACL-HLT} 2019, Minneapolis, MN, USA, June 2-7, 2019, Volume
  1 (Long and Short Papers)}}, \bibinfo{publisher}{Association for
  Computational Linguistics}, pp. \bibinfo{pages}{4171--4186},
  \doi{10.18653/v1/n19-1423}.

\bibitemdeclare{incollection}{wordnet}
\bibitem{wordnet}
\bibinfo{author}{Christiane \surnamestart Fellbaum\surnameend}
  (\bibinfo{year}{2010}): \emph{\bibinfo{title}{WordNet}}.
\newblock In: {\sl \bibinfo{booktitle}{Theory and applications of ontology:
  computer applications}}, \bibinfo{publisher}{Springer}, pp.
  \bibinfo{pages}{231--243}, \doi{10.1093/ijl/17.2.161}.

\bibitemdeclare{article}{DBLP:journals/tse/GansnerKNV93}
\bibitem{DBLP:journals/tse/GansnerKNV93}
\bibinfo{author}{Emden~R. \surnamestart Gansner\surnameend},
  \bibinfo{author}{Eleftherios \surnamestart Koutsofios\surnameend},
  \bibinfo{author}{Stephen~C. \surnamestart North\surnameend} \&
  \bibinfo{author}{Kiem{-}Phong \surnamestart Vo\surnameend}
  (\bibinfo{year}{1993}): \emph{\bibinfo{title}{A Technique for Drawing
  Directed Graphs}}.
\newblock {\sl \bibinfo{journal}{{IEEE} Trans. Software Eng.}}
  \bibinfo{volume}{19}(\bibinfo{number}{3}), pp. \bibinfo{pages}{214--230},
  \doi{10.1109/32.221135}.

\bibitemdeclare{inproceedings}{DBLP:conf/icms/GinevIJKKOSSTW16}
\bibitem{DBLP:conf/icms/GinevIJKKOSSTW16}
\bibinfo{author}{Deyan \surnamestart Ginev\surnameend}, \bibinfo{author}{Mihnea
  \surnamestart Iancu\surnameend}, \bibinfo{author}{Constantin \surnamestart
  Jucovschi\surnameend}, \bibinfo{author}{Andrea \surnamestart
  Kohlhase\surnameend}, \bibinfo{author}{Michael \surnamestart
  Kohlhase\surnameend}, \bibinfo{author}{Akbar \surnamestart
  Oripov\surnameend}, \bibinfo{author}{J{\"{u}}rgen \surnamestart
  Schefter\surnameend}, \bibinfo{author}{Wolfram \surnamestart
  Sperber\surnameend}, \bibinfo{author}{Olaf \surnamestart Teschke\surnameend}
  \& \bibinfo{author}{Tom \surnamestart Wiesing\surnameend}
  (\bibinfo{year}{2016}): \emph{\bibinfo{title}{The SMGloM Project and System:
  Towards a Terminology and Ontology for Mathematics}}.
\newblock In \bibinfo{editor}{Gert{-}Martin \surnamestart Greuel\surnameend},
  \bibinfo{editor}{Thorsten \surnamestart Koch\surnameend},
  \bibinfo{editor}{Peter \surnamestart Paule\surnameend} \&
  \bibinfo{editor}{Andrew~J. \surnamestart Sommese\surnameend}, editors: {\sl
  \bibinfo{booktitle}{Mathematical Software - {ICMS} 2016 - 5th International
  Conference, Berlin, Germany, July 11-14, 2016, Proceedings}}, {\sl
  \bibinfo{series}{Lecture Notes in Computer Science}} \bibinfo{volume}{9725},
  \bibinfo{publisher}{Springer}, pp. \bibinfo{pages}{451--457},
  \doi{10.1007/978-3-319-42432-3\_58}.

\bibitemdeclare{article}{Deyan1}
\bibitem{Deyan1}
\bibinfo{author}{Deyan \surnamestart Ginev\surnameend} \&
  \bibinfo{author}{Bruce~R. \surnamestart Miller\surnameend}
  (\bibinfo{year}{2019}): \emph{\bibinfo{title}{Scientific Statement
  Classification over arXiv.org}}.
\newblock {\sl \bibinfo{journal}{CoRR}} \bibinfo{volume}{abs/1908.10993}.
\newblock \urlprefix\url{http://arxiv.org/abs/1908.10993}.

\bibitemdeclare{inproceedings}{hearst-1992-automatic}
\bibitem{hearst-1992-automatic}
\bibinfo{author}{Marti~A. \surnamestart Hearst\surnameend}
  (\bibinfo{year}{1992}): \emph{\bibinfo{title}{Automatic Acquisition of
  Hyponyms from Large Text Corpora}}.
\newblock In: {\sl \bibinfo{booktitle}{{COLING} 1992 Volume 2: The 14th
  {I}nternational {C}onference on {C}omputational {L}inguistics}},
  \doi{10.3115/992133.992154}.

\bibitemdeclare{article}{DBLP:journals/corr/HuangXY15}
\bibitem{DBLP:journals/corr/HuangXY15}
\bibinfo{author}{Zhiheng \surnamestart Huang\surnameend}, \bibinfo{author}{Wei
  \surnamestart Xu\surnameend} \& \bibinfo{author}{Kai \surnamestart
  Yu\surnameend} (\bibinfo{year}{2015}): \emph{\bibinfo{title}{Bidirectional
  {LSTM-CRF} Models for Sequence Tagging}}.
\newblock {\sl \bibinfo{journal}{CoRR}} \bibinfo{volume}{abs/1508.01991}.
\newblock \urlprefix\url{http://arxiv.org/abs/1508.01991}.

\bibitemdeclare{inproceedings}{DBLP:conf/cikm/KaliszykKMR16}
\bibitem{DBLP:conf/cikm/KaliszykKMR16}
\bibinfo{author}{Cezary \surnamestart Kaliszyk\surnameend},
  \bibinfo{author}{Michael \surnamestart Kohlhase\surnameend},
  \bibinfo{author}{Dennis \surnamestart M{\"{u}}ller\surnameend} \&
  \bibinfo{author}{Florian \surnamestart Rabe\surnameend}
  (\bibinfo{year}{2016}): \emph{\bibinfo{title}{A Standard for Aligning
  Mathematical Concepts}}.
\newblock In \bibinfo{editor}{Andrea \surnamestart Kohlhase\surnameend},
  \bibinfo{editor}{Paul \surnamestart Libbrecht\surnameend},
  \bibinfo{editor}{Bruce~R. \surnamestart Miller\surnameend},
  \bibinfo{editor}{Adam \surnamestart Naumowicz\surnameend},
  \bibinfo{editor}{Walther \surnamestart Neuper\surnameend},
  \bibinfo{editor}{Pedro \surnamestart Quaresma\surnameend},
  \bibinfo{editor}{Frank~Wm. \surnamestart Tompa\surnameend} \&
  \bibinfo{editor}{Martin \surnamestart Suda\surnameend}, editors: {\sl
  \bibinfo{booktitle}{Joint Proceedings of the FM4M, MathUI, and ThEdu
  Workshops, Doctoral Program, and Work in Progress at the Conference on
  Intelligent Computer Mathematics 2016 co-located with the 9th Conference on
  Intelligent Computer Mathematics {(CICM} 2016), Bialystok, Poland, July
  25-29, 2016}}, {\sl \bibinfo{series}{{CEUR} Workshop Proceedings}}
  \bibinfo{volume}{1785}, \bibinfo{publisher}{CEUR-WS.org}, pp.
  \bibinfo{pages}{229--244}.
\newblock \urlprefix\url{http://ceur-ws.org/Vol-1785/W24.pdf}.

\bibitemdeclare{article}{word2vec}
\bibitem{word2vec}
\bibinfo{author}{Tom{\'{a}}s \surnamestart Mikolov\surnameend},
  \bibinfo{author}{Ilya \surnamestart Sutskever\surnameend},
  \bibinfo{author}{Kai \surnamestart Chen\surnameend}, \bibinfo{author}{Greg
  \surnamestart Corrado\surnameend} \& \bibinfo{author}{Jeffrey \surnamestart
  Dean\surnameend} (\bibinfo{year}{2013}): \emph{\bibinfo{title}{Distributed
  Representations of Words and Phrases and their Compositionality}}.
\newblock {\sl \bibinfo{journal}{CoRR}} \bibinfo{volume}{abs/1310.4546},
  \doi{10.5555/2999792.2999959}.

\bibitemdeclare{inproceedings}{DBLP:conf/nips/NickelK17}
\bibitem{DBLP:conf/nips/NickelK17}
\bibinfo{author}{Maximilian \surnamestart Nickel\surnameend} \&
  \bibinfo{author}{Douwe \surnamestart Kiela\surnameend}
  (\bibinfo{year}{2017}): \emph{\bibinfo{title}{Poincar{\'{e}} Embeddings for
  Learning Hierarchical Representations}}.
\newblock In \bibinfo{editor}{Isabelle \surnamestart Guyon\surnameend},
  \bibinfo{editor}{Ulrike \surnamestart von Luxburg\surnameend},
  \bibinfo{editor}{Samy \surnamestart Bengio\surnameend},
  \bibinfo{editor}{Hanna~M. \surnamestart Wallach\surnameend},
  \bibinfo{editor}{Rob \surnamestart Fergus\surnameend},
  \bibinfo{editor}{S.~V.~N. \surnamestart Vishwanathan\surnameend} \&
  \bibinfo{editor}{Roman \surnamestart Garnett\surnameend}, editors: {\sl
  \bibinfo{booktitle}{Advances in Neural Information Processing Systems 30:
  Annual Conference on Neural Information Processing Systems 2017, December
  4-9, 2017, Long Beach, CA, {USA}}}, pp. \bibinfo{pages}{6338--6347}.
\newblock
  \urlprefix\url{https://proceedings.neurips.cc/paper/2017/hash/59dfa2df42d9e3d41f5b02bfc32229dd-Abstract.html}.

\bibitemdeclare{article}{DBLP:journals/jmlr/PedregosaVGMTGBPWDVPCBPD11}
\bibitem{DBLP:journals/jmlr/PedregosaVGMTGBPWDVPCBPD11}
\bibinfo{author}{Fabian \surnamestart Pedregosa\surnameend},
  \bibinfo{author}{Ga{\"{e}}l \surnamestart Varoquaux\surnameend},
  \bibinfo{author}{Alexandre \surnamestart Gramfort\surnameend},
  \bibinfo{author}{Vincent \surnamestart Michel\surnameend},
  \bibinfo{author}{Bertrand \surnamestart Thirion\surnameend},
  \bibinfo{author}{Olivier \surnamestart Grisel\surnameend},
  \bibinfo{author}{Mathieu \surnamestart Blondel\surnameend},
  \bibinfo{author}{Peter \surnamestart Prettenhofer\surnameend},
  \bibinfo{author}{Ron \surnamestart Weiss\surnameend},
  \bibinfo{author}{Vincent \surnamestart Dubourg\surnameend},
  \bibinfo{author}{Jake \surnamestart VanderPlas\surnameend},
  \bibinfo{author}{Alexandre \surnamestart Passos\surnameend},
  \bibinfo{author}{David \surnamestart Cournapeau\surnameend},
  \bibinfo{author}{Matthieu \surnamestart Brucher\surnameend},
  \bibinfo{author}{Matthieu \surnamestart Perrot\surnameend} \&
  \bibinfo{author}{Edouard \surnamestart Duchesnay\surnameend}
  (\bibinfo{year}{2011}): \emph{\bibinfo{title}{Scikit-learn: Machine Learning
  in Python}}.
\newblock {\sl \bibinfo{journal}{J. Mach. Learn. Res.}} \bibinfo{volume}{12},
  pp. \bibinfo{pages}{2825--2830}.
\newblock \urlprefix\url{http://dl.acm.org/citation.cfm?id=2078195}.

\bibitemdeclare{inproceedings}{DBLP:conf/emnlp/PenningtonSM14}
\bibitem{DBLP:conf/emnlp/PenningtonSM14}
\bibinfo{author}{Jeffrey \surnamestart Pennington\surnameend},
  \bibinfo{author}{Richard \surnamestart Socher\surnameend} \&
  \bibinfo{author}{Christopher~D. \surnamestart Manning\surnameend}
  (\bibinfo{year}{2014}): \emph{\bibinfo{title}{Glove: Global Vectors for Word
  Representation}}.
\newblock In \bibinfo{editor}{Alessandro \surnamestart Moschitti\surnameend},
  \bibinfo{editor}{Bo~\surnamestart Pang\surnameend} \& \bibinfo{editor}{Walter
  \surnamestart Daelemans\surnameend}, editors: {\sl
  \bibinfo{booktitle}{Proceedings of the 2014 Conference on Empirical Methods
  in Natural Language Processing, {EMNLP} 2014, October 25-29, 2014, Doha,
  Qatar, {A} meeting of SIGDAT, a Special Interest Group of the {ACL}}},
  \bibinfo{publisher}{{ACL}}, pp. \bibinfo{pages}{1532--1543},
  \doi{10.3115/v1/d14-1162}.

\bibitemdeclare{inproceedings}{DBLP:conf/iclr/TifreaBG19}
\bibitem{DBLP:conf/iclr/TifreaBG19}
\bibinfo{author}{Alexandru \surnamestart Tifrea\surnameend},
  \bibinfo{author}{Gary \surnamestart B{\'{e}}cigneul\surnameend} \&
  \bibinfo{author}{Octavian{-}Eugen \surnamestart Ganea\surnameend}
  (\bibinfo{year}{2019}): \emph{\bibinfo{title}{Poincare Glove: Hyperbolic Word
  Embeddings}}.
\newblock In: {\sl \bibinfo{booktitle}{7th International Conference on Learning
  Representations, {ICLR} 2019, New Orleans, LA, USA, May 6-9, 2019}},
  \bibinfo{publisher}{OpenReview.net}.
\newblock \urlprefix\url{https://openreview.net/forum?id=Ske5r3AqK7}.

\bibitemdeclare{inproceedings}{wang-etal-2017-short}
\bibitem{wang-etal-2017-short}
\bibinfo{author}{Chengyu \surnamestart Wang\surnameend},
  \bibinfo{author}{Xiaofeng \surnamestart He\surnameend} \&
  \bibinfo{author}{Aoying \surnamestart Zhou\surnameend}
  (\bibinfo{year}{2017}): \emph{\bibinfo{title}{A Short Survey on Taxonomy
  Learning from Text Corpora: Issues, Resources and Recent Advances}}.
\newblock In: {\sl \bibinfo{booktitle}{Proceedings of the 2017 Conference on
  Empirical Methods in Natural Language Processing}},
  \bibinfo{publisher}{Association for Computational Linguistics},
  \bibinfo{address}{Copenhagen, Denmark}, pp. \bibinfo{pages}{1190--1203},
  \doi{10.18653/v1/D17-1123}.

\end{thebibliography}
\end{document}